\documentclass[a4paper,11pt]{article}
\pdfoutput=1 

\usepackage{jinstpub} 

\title{\boldmath BBU instability in rectangular dielectric resonator}

 \author[1]{G. V. Sotnikov,\note{Corresponding author.}}
 \author{K. V. Galaydych,}
 \author{R. R. Kniaziev,}
 \author{and I. N. Onishchenko}
 \affiliation{NSC ''Kharkov Institute of Physics and Technology'',\\ 1 Akademicheskaya Str., 61108 Kharkov, Ukraine}

\emailAdd{sotnikov@kipt.kharkov.ua,gennadiy.sotnikov@gmail.com}

\abstract{Studies were made into the arise and an evolution of the beam breakup (BBU) instability in a rectangular dielectric resonator under excitation by a sequence of relativistic electron bunches. The dielectric resonator is a metal rectangular waveguide $R_{26}$ $(45\,mm\times 90\,mm)$ with Teflon dielectric slabs $8.2\,mm$ thick (dielectric constant $\varepsilon=2.051$) located along the wide side of the resonator. The wavelength of the $LM_{21}$ operating mode having a symmetric profile of the longitudinal electric field component is $53.2\,mm$. The electron energy of bunches is $4.5\,MeV$ , the charge of each bunch is $6.4\,nC$, the bunch repetition period is equal to twice the wavelength of the $LM_{21}$ mode. By the use of numerical PIC simulations, the charge losses of electron bunches on the dielectric plates were investigated as the bunches were displaced relative to the cavity axis. It is found that the charge losses on the dielectric slabs due to the BBU instability do not exceed $5\%$. When the bunch repetition period is changed to a multiple of another eigenfrequency (e.g., the $LM_{11}$ mode), the charge losses of drive bunches do not change appreciably.}


\proceeding{XIII$^{\text{th}}$ International Symposium «Radiation from Relativistic Electrons in Periodic Structures»\\
  September 16-20, 2019\\
  Belgorod, Russian Federation}

\begin{document}
\maketitle
\flushbottom

\section{Introduction}
\label{sec:intro}

One of the advanced charged particles acceleration methods is the acceleration by the wakefields, which are excited by a single relativistic electron bunch or a bunch train in dielectric-loaded structures.
%
Recently more attention has been paid to the studies of planar and rectangular accelerating structures (the so-called multizone dielectric structures) with dielectric slabs and vacuum channels for charged particles~\cite{Tremaine1997PRE,Zha1997PRE,Park2001PP,Park2002AAC-10,
Mar2001PRSTAB,Xiao2002AAC-10,Mar2002AAC-10,%
Wang2004PRSTAB,Wang2006PRSTAB,Xiao2001PRE,Mih2012PRSTAB}. These structures have a number of technological advantages over cylindrical ones. Among them is the possibility that a working mode in the rectangular dielectric accelerating structures can be selected with a symmetrical distribution of the axial electric field in the transverse direction inside the vacuum channel. As a consequence, this reduces the transverse forces acting on both the drive bunches and the test bunch, thereby improving not only the efficiency of the accelerating structure excitation, but also the test beam acceleration.

One of the key issues of the accelerators development is the transverse stability of bunches. In the case of dielectric wakefield accelerators, this issue is especially important, because the accelerating field is created here by the drive bunches, rather than by an external RF source, as in conventional accelerators. As a consequence of the instability, aside from the bunch parameters degradation (e.g. increase in the transverse size and energy spread, etc.), it is also possible that particles may deposit on the dielectric surface. In turn, this would lead to changes in the dielectric material parameters and, as a consequence, to detuning of the Cherenkov resonance conditions. The important case of transverse instability is the asymmetric bunch injection. So far, theoretical studies of wakefield excitation in the dielectric structures under asymmetric electron bunch injection have been performed in the waveguide problem statement~\cite{Li2014PRSTAB,Gai1997PRE,Baturin2018PRSTAB,Altmark_2012}, without taking into account the finite length of the accelerating module. In the resonator, unlike the waveguide, all the bunches of regular sequence are involved in the total electromagnetic field formation and in the accumulation of electromagnetic field energy from a large number of bunches that are regularly injected, this being the most effective use of dielectric resonators~\cite{Bal2003TPL,Mar2001AAC-9}. Therefore, the aspects acting on the charged particles transverse dynamics, require a detailed theoretical analysis for further development of an efficient creation of the dielectric accelerating resonator structures. The foregoing determines the relevance and necessity of the presented theoretical studies.

\section{Statement of the problem}
\begin{figure}[htbp]
\centering 
\includegraphics[width=\textwidth]{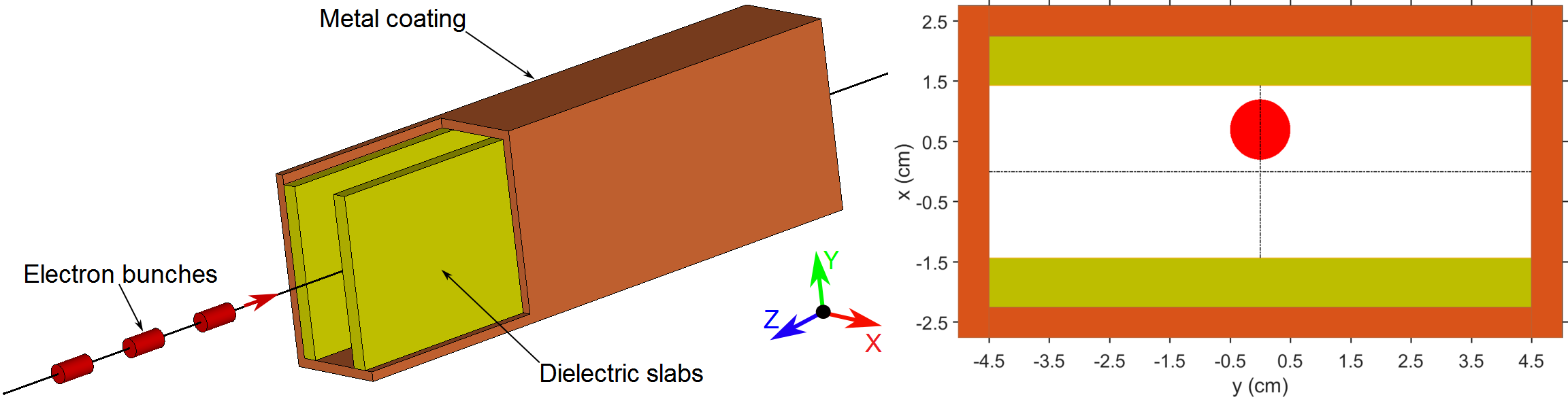}\hfil
\caption{\label{fig:01} General view (left) and side view (right) of the three-zone dielectric rectangular resonator. The metal coating (orange), dielectric slabs (yellow), and electron bunches (red) are shown schematically. The electron bunches, externally injected with an offset, move 
in parallel to the resonator axis. 
}
\end{figure}

The three-zone dielectric resonator under investigation is a rectangular metal resonator with dielectric slabs placed in parallel to one of the walls. In the vacuum channel a train of relativistic electron bunches (bunch repetition frequency (BRF) is $f_m$) is injected with an offset into the resonator in parallel to the resonator axis (as it is presented in figure~\ref{fig:01}).

The main goal of the present paper was to analyze the dynamics of bunch-excited wakefields and the dynamics of the drive electron bunches, which excite these wakefields in the case of their off-axis injection. We assume that the end walls of the resonator are closed by metal grids transparent for charged particles and nontransparent for the excited electromagnetic field. The simulations were carried out with the parameters similar to those in the experiments expected to be done at Kharkiv Institute of Physics and Technology; they are presented in table~\ref{tab:01}.
\begin{table}[htbp]
\centering
\caption{\label{tab:01} Parameters of the three-zone dielectric resonator and the electron bunch train.}
\smallskip
\begin{tabular}{|l|r|c|l|r|c|}
\hline
\textbf{Parameter }& \textbf{Value} &\textbf{ Unit}& Parameter & \textbf{Value} &\textbf{ Unit} \\
\hline
Resonator width     & 4.5  & cm & Bunch charge & 6.4 & nC\\
Resonator height    & 9.0  & cm & Energy of bunch electrons & 4.5 & MeV\\
Slab thickness      & 0.82 & cm & Bunch RMS radius $\sigma_r$ & 0.25 & cm\\
Drift channel width & 2.86 & cm & Bunch RMS length $\sigma_z$ & 0.47 & cm\\
Dielectric constant & 2.051&    & BRF $f_m$ ($LSM_{2,1,12})$ & 5.606 & GHz\\               Resonator length ($5.606\,GHz$) & 31.92  & cm & BRF $f_m$ ($LSM_{1,1,12})$ & 3.324 & GHz\\
Resonator length ($3.324\,GHz$)    & 52.83  & cm & Bunch offset & 0.9 & cm\\

\hline
\end{tabular}
\end{table}
The resonator length has been chosen such that it should be equal to six wavelengths of the $LSM_{2,1,12}$ and $LSM_{1,1,12}$ modes, resonant with a bunch\footnote{It should be noted that in comparison with~\cite{Xiao2001PRE,Mih2012PRSTAB} we keep the general classification of modes given in ref.~\cite{Wang2006PRSTAB}, where the modes having  symmetric and asymmetric transverse distributions of longitudinal electric field are described by one combined equation.}. These modes have the asymmetric ($LSM_{2,1,12}$) and symmetric ($LSM_{1,1,12}$) transverse distributions of the transverse forces, which act on the test particle in the vacuum channel. This choice of the resonator parameters provided realization of the conditions of the resonator wakefield accelerator concept~\cite{Onish2008TP}. For the particle-in-cell simulations the CST Particle Studio was used~\cite{CST-PS}.

\section{Simulations results}
\begin{figure}[htbp]
\centering 
\includegraphics[width=\textwidth]{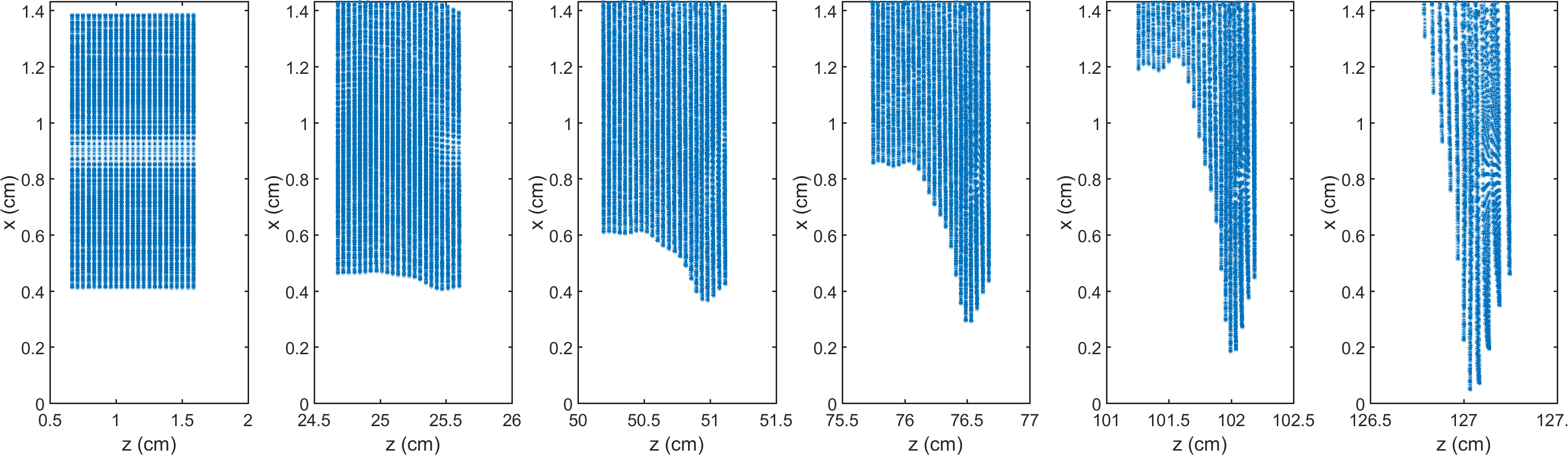}\hfil
\caption{\label{fig:02} Dynamics of the off-axis single bunch on the $x-z$ plane. Vertical axis limits correspond to half of vacuum channel width, horizontal axis limits correspond to the frame around the bunch. The initial horizontal offset of the bunch is $0.9\,cm$, the length of the resonator is $127.67\,cm$.}
\end{figure}

We start with the case of wakefield excitation by a single relativistic electron bunch. In order for the effect of the bunch offset to be more pronounced, we injected it near the dielectric slab and also increased the resonator length by four times (compared to the  planned in the experiment). Figure~\ref{fig:02} demonstrates the simulation results as the snapshots of bunch particles positions on the $x-z$ plane at consecutive moments of time. The first moment (leftmost shot) corresponds to the case that the bunch is fully injected into the vacuum channel of the resonator. The last moment (rightmost shot) shows to that the bunch has reached the resonator output.

It can be seen, that the transverse displacement of the tail particles is greater than that of the head ones. This dynamics of the bunch is the result of the initial offset. Consider also the fact that the amplitude of the transverse component of the excited wakefield is zero in front of the bunch and it grows linearly from head to tail of the bunch. In the absence of the initial offset there is no bunch displacement as a whole. In the on-axis injection case the particle dynamics is symmetrical and will be demonstrated below. Note that the bunch losses in the tail are greater than in the bunch head. Thus, we can conclude, that the BBU instability for the case of a single bunch in the rectangular dielectric resonator does exist and leads to undesirable electron bunch degradation up to the total beam loss.

\begin{figure}[htbp]
\centering 
\includegraphics[width=\textwidth]{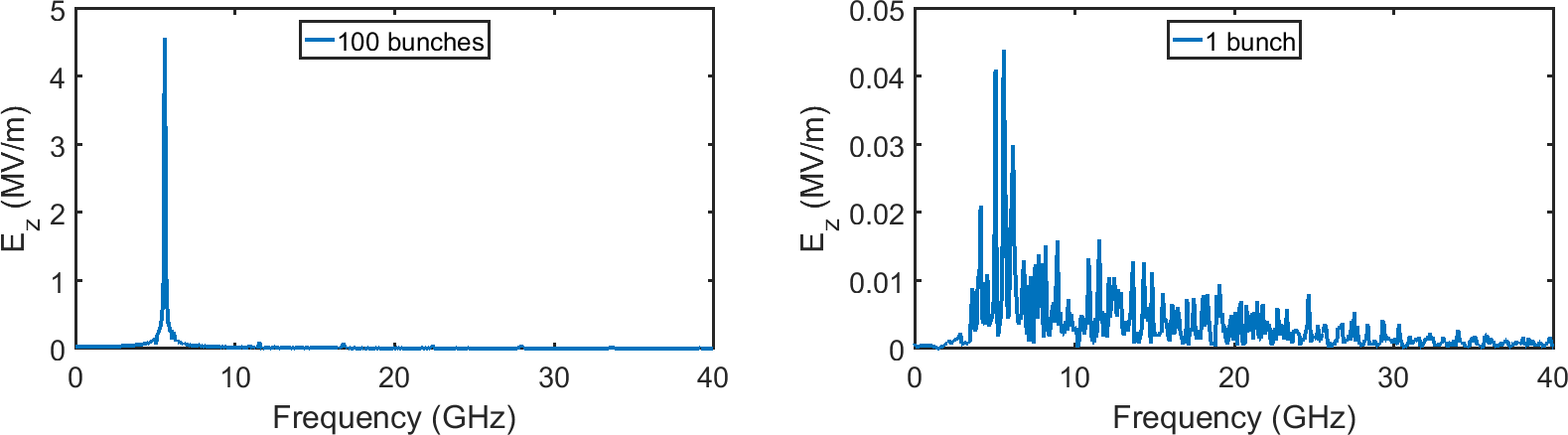}\hfil
\caption{\label{fig:03} The spectra of the longitudinal electric field (at the point $x = 0.9\,cm,\,y = 0,\,z = 15.96\,cm$ ), excited by a sequence of $100$ bunches with $f_m=5.606\,GHz$ (left), and 
by the single electron bunch (right).}
\end{figure}

It is well known that the parasitic modes excitation (e.g. dipole mode) accounts for the increase in the transverse deflecting force, which causes the BBU occurrence in the accelerating structures. Therefore spectral analysis of the bunch-excited electromagnetic field was carried out for the case of multibunch resonator excitation. The results of the spectral analysis are presented in figure~\ref{fig:03}.

Figure~\ref{fig:03} demonstrates that in the case of single electron bunch excitation, the spectrum of the electromagnetic field exhibits many eigenfrequencies of the dielectric resonator (figure~\ref{fig:03}, right plot). The wide frequency spectrum is the result of the excitation of a large number of nonresonant eigenfrequencies, which can be interpreted as a field of transient radiation, generated by the electron bunch during its injection into the resonator and output from it. As one can see, the spectrum of the $E_z$ field in the case of the bunch sequence excitation is represented by a single frequency. The maximum in the spectrum corresponds to the bunch repetition frequency $f_m$, and there are no excited parasitic modes. This is due to the fact that the bunch sequence enhances the resonant mode as well as the modes the frequencies of which are close to multiple resonant frequencies, simultaneously suppressing the nonresonant modes.

\begin{figure}[htbp]
\centering 
\includegraphics[width=\textwidth]{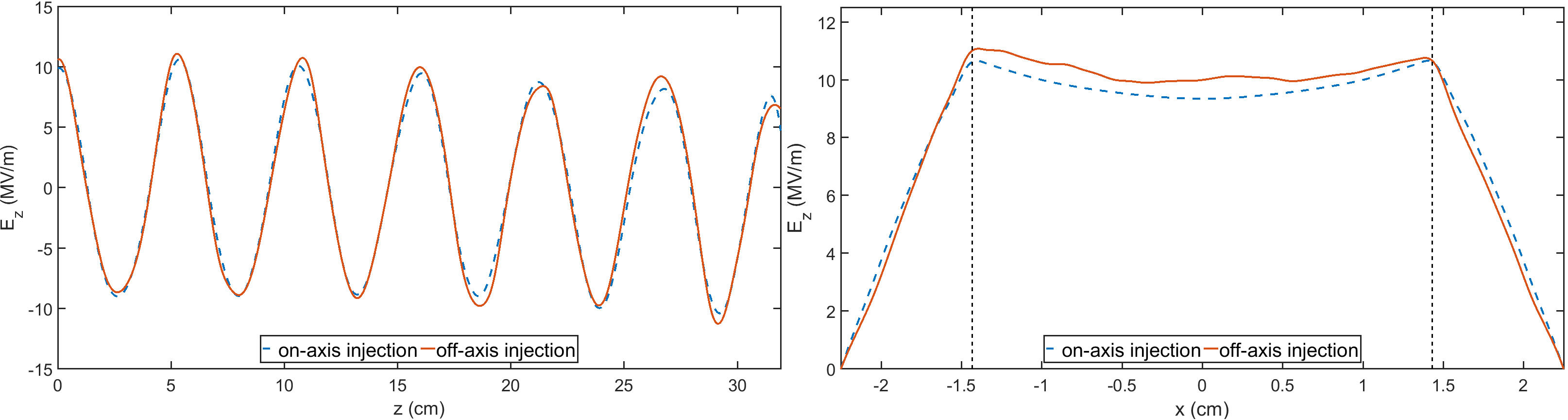}\hfil
\caption{\label{fig:04} Longitudinal (left) and transverse (right) distributions of the axial component of the electric field excited by a sequence of $100$ electron bunches injected at $f_m=5.606\,GHz$. The vertical dashed lines indicate the positions of dielectric slabs. The longitudinal distribution is calculated at the shifted bunch center (at  $x = 0.9\,cm,\,y = 0$), the transversal distribution is calculated at the maximum of the axial electric field (at $y = 0,\,z = 27.93\,cm$  ). After passage of $100$ bunches practicaly only one resonant $LSM_{2,1,12}$ mode survives.}
\end{figure}

The off-axis bunch injection may lead to the distortion of the bunch-excited field components, which, in turn, have an effect on drive and witness bunches. The distributions of the axial and transverse components of the electric and magnetic fields, both in the longitudinal ($z$) and in the transverse ($x$) directions, were obtained and analyzed as well. Figure~\ref{fig:04} shows the axial and transverse distributions of the longitudinal component of the electric field, excited in the resonator by a sequence of $100$ electron bunches for the cases of the on-and-off axis injection. The presented spatial distributions of the field components correspond to the moment of time, when the last bunch of the sequence reached the resonator output. As the number of drive electron bunches injected into the resonator increases, the amplitude of the electric field in the resonator also increases, and in the longitudinal and transverse directions it becomes more monotonic with characteristic spatial period of the resonant mode. In this case, the electromagnetic field changes in time so that the drive bunches remain in the decelerating phases of the field, continuing to transfer their energy to the electromagnetic field. The increase in the amplitude of the electric field longitudinal component at off-axis injection is associated with the increase in the coupling between the resonance mode and the bunches (the resonance mode amplitude is maximum at the dielectric surface and minimum at the resonator axis).

\begin{figure}[htbp]
\centering 
\includegraphics[width=\textwidth]{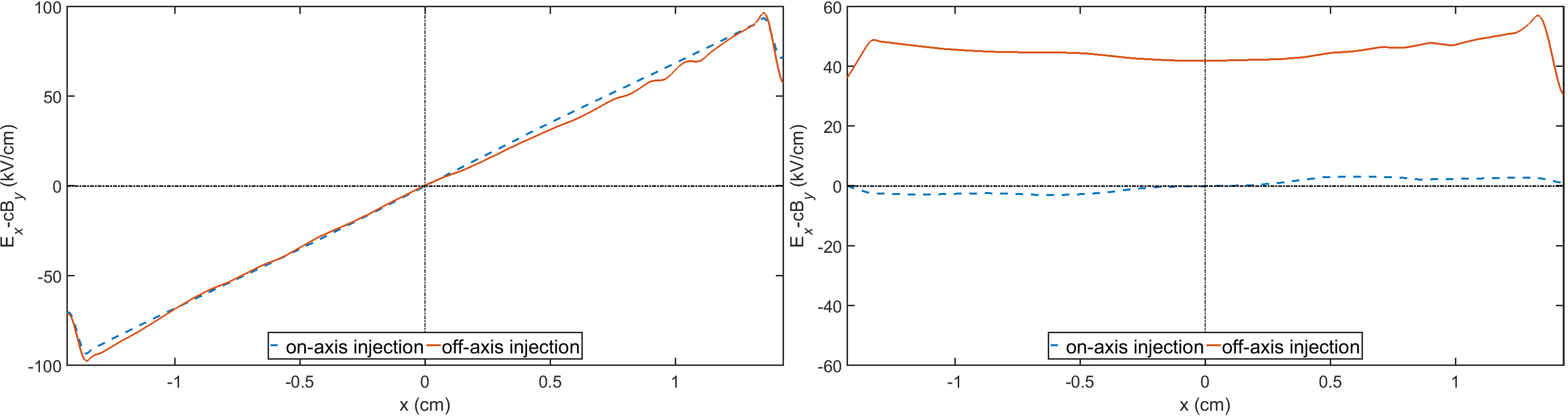}\hfil
\caption{\label{fig:05} The distributions of transverse forces, which act on the test particle across the resonator vacuum channel for the cases of: (left) $f_m=5.606\,GHz$ (at $y = 0,\,z = 27.93\,cm$ ) and (right) $f_m=3.324\,GHz$ (at $y = 0,\,z = 47.09\,cm$ ). The resonator is excited by a sequence of $100$ electron bunches.}
\end{figure}

Figure~\ref{fig:05} presents the distribution of the transverse force acting on the test particle across the vacuum channel, for the cases of on-and-off axis injection at repetition rate equal to the frequencies of resonant modes $LSM_{2,1,12}$ and $LSM_{1,1,12}$, respectively.

It can be seen that: (i) even at significant bunch offset ($0.9\,cm$) the spatial distribution of the field components remain almost the same (in comparison with the on-axis injection); (ii) the transverse profiles are similar to the profiles of resonant $LSM_{2,1,12}$ or $LSM_{1,1,12}$ modes consequently. A slight distortion of the profiles of the electromagnetic field components is associated with the excitation of the modes at frequencies that are multiples of the repetition rate $f_m$ of the drive bunches. These results are completely consistent with the data of the spectral analysis shown in figure~\ref{fig:03}. It should be noted that for the case of the resonant $LSM_{1,1,12}$ mode the transverse force distributions differ significantly for the on-axis and off-axis bunch injection. This difference is due to the fact that the $LSM_{1,1,12}$ mode has the asymmetric transversal distribution of the axial electric field in the vacuum channel cross section that leads to weak coupling between the on-axis bunch and the field of this mode, in contrast with the off-axis case.

\begin{figure}[htbp]
\centering 
\includegraphics[width=\textwidth]{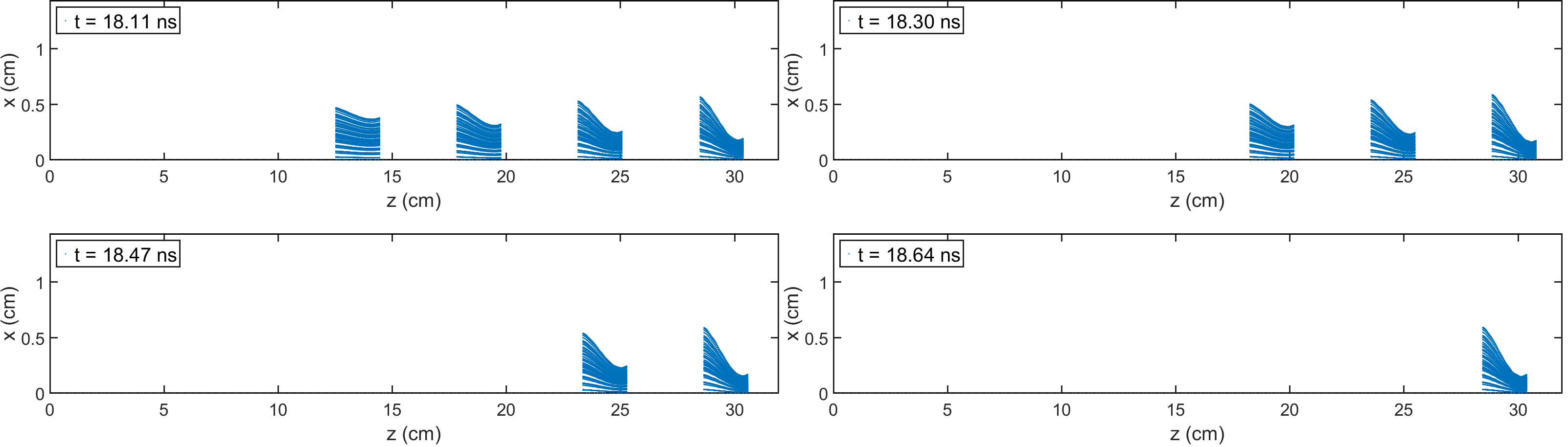}
\caption{\label{fig:06} Bunch particle dynamics in the $x-z$ plane in the case of injection along the axis of the resonator for four consecutive time moments after injection of the last bunch of the sequence. Bunch repetition frequency is $5.606\,GHz$. Due to the on-axis injection, and, hence, the symmetrical particle dynamics, only half of the particles are shown.}
\end{figure}

As we are interested in the dynamics of drive bunches injected off-axis, detailed studies were made for both the transverse and longitudinal directions. Comparison with the case of on-axis injection was carried out, too. To gain a better understanding of the bunch dynamics, figure~\ref{fig:06} shows the particles positions in the vacuum channel for the consecutive time moments. The positions of particles are given in the $x-z$ plane.

\begin{figure}[htbp]
\centering 
\includegraphics[width=\textwidth]{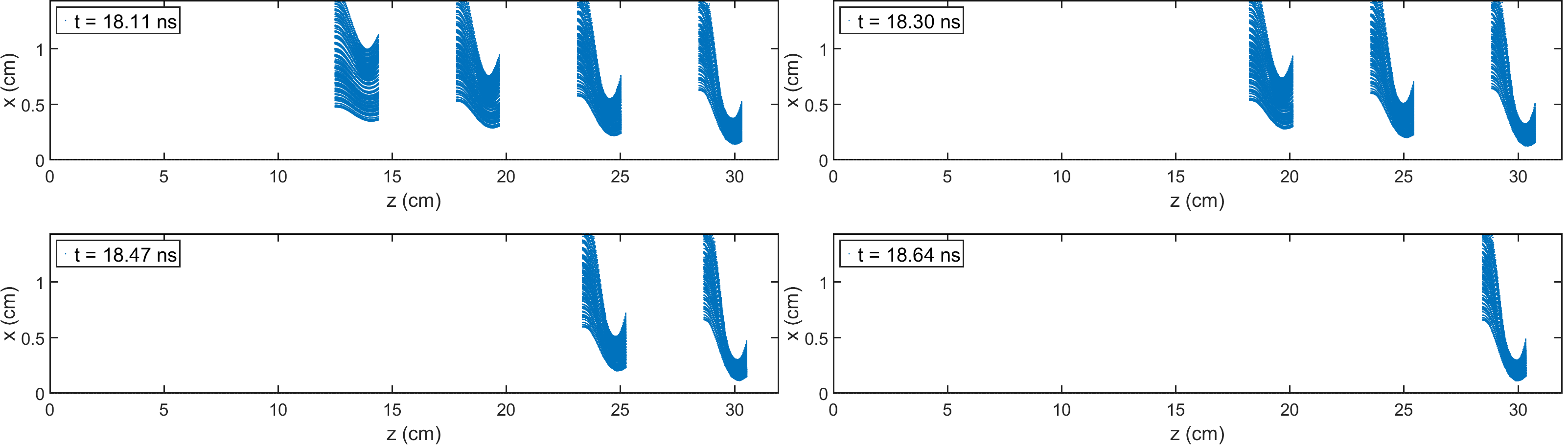}
\caption{\label{fig:07} Bunch particles dynamics in the $x-z$ plane in the case of asymmetric injection for the four consecutive time moments after injection of the last bunch of the sequence.}
\end{figure}

The first time moment corresponds to the moment when the last bunch of the sequence is fully injected into the resonator and has passed half the resonator length, the last time moment -- when this bunch reached the output of the resonator. The characteristic feature of the symmetric resonant mode is the simultaneous focusing in one transverse direction, and defocusing in the other one. In which of the two transverse directions is focusing, and in which is defocusing, depends on the phase of the particle relative to the resonant mode field. It is seen in figure~\ref{fig:06} shows that as the bunches pass along the resonator vacuum channel, a part of each bunch is focused in the $x-z$ plane, and the other part of the same bunch is defocused in the same plane. The transverse dynamics of particles in the $y-z$ plane is directly opposite. The particles that defocus, in the $x-z$ plane get focused. The corresponding dynamics of the bunches in the $x-z$ plane in the case of off-axis injection is shown in figure~\ref{fig:07}.

There are no qualitative differences in the bunch dynamics in the $y-z$ plane (and therefore omitted in the paper) compared to the on-axis injection. Significant qualitative differences take place in the $x-z$ plane only. This is determined by the structure of the transverse components of the bunch-excited field. The bunch sequence enhances the resonant mode with asymmetric transverse field components distribution, and suppresses other modes, and, as a result, a significant part of each bunch backs on the resonator axis. Figure~\ref{fig:07} demonstrates that the charge losses occur for the bunch particles located at its periphery, therefore, the bunch charge losses should not be significant.

\begin{figure}[htbp]
\centering 
\includegraphics[width=\textwidth]{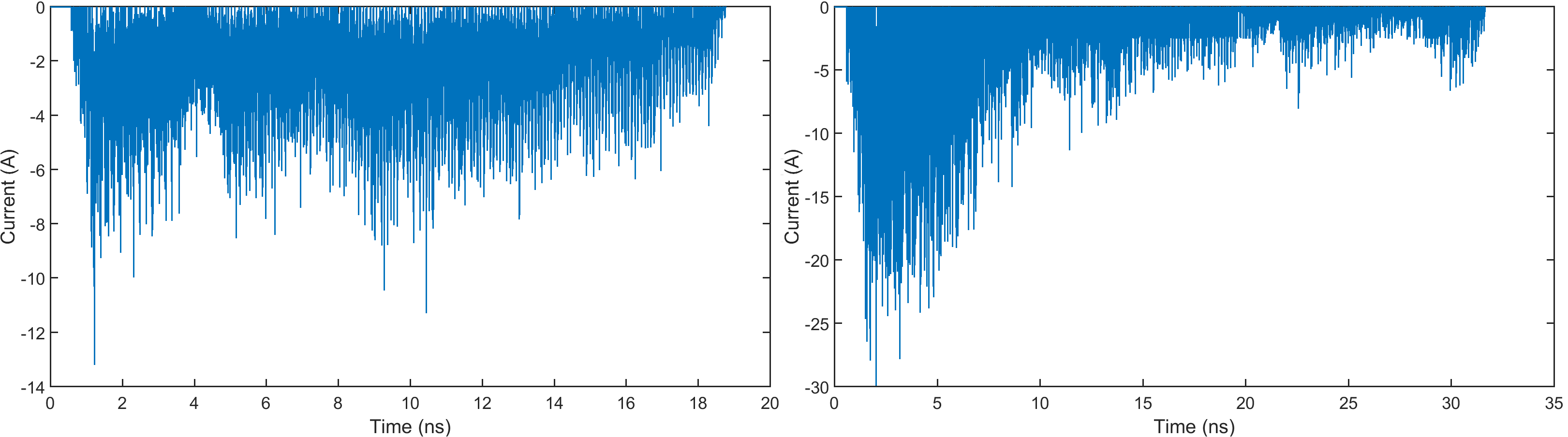}
\caption{\label{fig:08} The bunch train current through the dielectric slab surface as a function of time: (left) $f_m=5.606\,GHz$ (symmetric resonant mode), (right)  $f_m=3.324\,GHz$ (asymmetric resonant mode). Maximum input current in the simulations is 169 A.}
\end{figure}

As a quantitative characteristic of charge losses due to the particle deposition on the dielectric, the current through the dielectric slab surface was chosen for the analysis. The bunch current losses as a function of time for the cases of symmetric and asymmetric resonant modes are presented in figure~\ref{fig:08}.

The analysis of the above presented currents leads to the conclusion that the current through the dielectric slab surface does not tend to increase with time. In turn, this means that a gradual increase of charge losses does not occur. It should be noted that the current (in the both cases) increases until the first bunch leaves the resonator, after that the current begins to decrease. It was found that the charge losses on the dielectric slabs due to the BBU instability do not exceed 5\%. It is well known that the wakefield excitation in a resonator does not differ from that in the semi-infinite waveguide, until the first bunch leaves the resonator. In turn, simulation of the wakefield excitation by the off-axis bunches in a corresponding semi-infinite structure has shown, that the initial bunch offset increases with increase in the distance travelled by these bunches. The main difference between these two cases is that in the waveguide case, the bunch-excited field is a travelling wave, and in the resonator case it is the standing wave. As all the bunches are injected in the decelerating phases (in order to increase the axial electric field), and the transverse field components of the resonant $LSM_{2,1,12}$ mode act as a focusing and back the bunches to the resonator axis. In the case of the $LSM_{1,1,12}$ resonant mode (in contrast to the resonant $LSM_{2,1,12}$ mode), drive bunches expand in the $x-z$ plane, but also do not deflect to the dielectric plate as a whole. This means that the longitudinal boundedness of the structure does not allow the BBU instability to arise.

\section{Conclusions}
In this paper we have presented detailed simulation studies of the BBU instability in the three-zone rectangular dielectric resonator, excited by a regular sequence of relativistic electron bunches. Fourier analysis of the electromagnetic field, excited in the resonator, has shown that the maximum in the spectrum corresponds to the resonant frequency of the bunch repetition, and there are no excited parasitic modes, which could lead to the BBU instability. It has been demonstrated that the presence of the bunch offset does not significantly affect the spatial distribution of the electromagnetic field components excited in the resonator. The analysis of the transverse dynamics of the electron bunches, as well as the current analysis, have demonstrated, that the increase in current losses through the dielectric slab does not occur with time. Thus, the initial bunch offset does not lead to the BBU instability, and is not the critical point, which requires high injection accuracy for the future development and operation of the resonant wakefield accelerator.

\acknowledgments

This work was supported by NAS of Ukraine, the program "Perspective researches on plasma physics, controlled thermonuclear fusion and plasma technologies" (Project P-1/63-2017), and by the Ukrainian budget program "Support for the most important directions of scientific researches" (K$\Pi$KBK 6541230).


\bibliographystyle{JHEP}
\bibliography{Bibliography}	



\end{document}